\documentclass[aps,prb,amsmath,amssymb,amsfont,showpacs,superscriptaddress,reprint]{revtex4-1}
\usepackage{amsmath, amssymb, amsfonts, multirow}
\usepackage{bm}
\usepackage[dvipsnames]{xcolor}
\usepackage{braket}
\usepackage[artemisia]{textgreek}  
\usepackage{float}
\usepackage[version=3]{mhchem}
\usepackage{tabularx}
\usepackage{graphicx}

\begin{document}

\title{Tuning Conductivity and Spin Dynamics in Few-Layer Graphene via \emph{In Situ} Potassium Exposure}

\author{B.~G.~M\'{a}rkus}
\affiliation{Department of Physics, Budapest University of Technology and Economics and MTA-BME Lend\"{u}let Spintronics Research Group (PROSPIN), P.O. Box 91, H-1521 Budapest, Hungary}

\author{O.~S\'agi}
\affiliation{Department of Physics, Budapest University of Technology and Economics and MTA-BME Lend\"{u}let Spintronics Research Group (PROSPIN), P.O. Box 91, H-1521 Budapest, Hungary}

\author{S.~Kollarics}
\affiliation{Department of Physics, Budapest University of Technology and Economics and MTA-BME Lend\"{u}let Spintronics Research Group (PROSPIN), P.O. Box 91, H-1521 Budapest, Hungary}

\author{K.~F.~Edelthalhammer}
\affiliation{Department of Chemistry and Pharmacy and Institute of Advanced Materials and Processes
(ZMP), University of Erlangen-N\"{u}rnberg, Nikolaus-Fiebiger-Strasse 10, 91058 Erlangen, Germany}

\author{A.~Hirsch}
\affiliation{Department of Chemistry and Pharmacy and Institute of Advanced Materials and Processes
(ZMP), University of Erlangen-N\"{u}rnberg, Nikolaus-Fiebiger-Strasse 10, 91058 Erlangen, Germany}

\author{F.~Hauke}
\affiliation{Department of Chemistry and Pharmacy and Institute of Advanced Materials and Processes
(ZMP), University of Erlangen-N\"{u}rnberg, Nikolaus-Fiebiger-Strasse 10, 91058 Erlangen, Germany}

\author{P.~Szirmai}
\affiliation{Institute of Physics of Complex Matter, FBS Swiss Federal Institute of Technology (EPFL), CH-1015 Lausanne, Switzerland}

\author{B.~N\'afr\'adi}
\affiliation{Institute of Physics of Complex Matter, FBS Swiss Federal Institute of Technology (EPFL), CH-1015 Lausanne, Switzerland}

\author{L.~Forr\'o}
\affiliation{Institute of Physics of Complex Matter, FBS Swiss Federal Institute of Technology (EPFL), CH-1015 Lausanne, Switzerland}

\author{F.~Simon}
\affiliation{Department of Physics, Budapest University of Technology and Economics and MTA-BME Lend\"{u}let Spintronics Research Group (PROSPIN), P.O. Box 91, H-1521 Budapest, Hungary}
\affiliation{Institute of Physics of Complex Matter, FBS Swiss Federal Institute of Technology (EPFL), CH-1015 Lausanne, Switzerland}


\begin{abstract}
Chemical modification, such as intercalation or doping of novel materials is of great importance for exploratory material science and applications in various fields of physics and chemistry. In the present work, we report the systematic intercalation of chemically exfoliated few-layer graphene with potassium while monitoring the sample resistance using microwave conductivity. We find that the conductivity of the samples increases by about an order of magnitude upon potassium exposure. The increased of number of charge carriers deduced from the ESR intensity also reflects this increment. The doped phases exhibit two asymmetric Dysonian lines in ESR, a usual sign of the presence of mobile charge carriers. The width of the broader component increases with the doping steps, however, the narrow components seem to have a constant line width.
\end{abstract}

\maketitle

\section{Introduction}

Several allotropes of carbon are well-known electron acceptor materials. The charge doping can be conveniently performed with the help of alkali and alkali earth metal atoms when these are brought into contact with the carbon. A good example of alkali doping of the zero dimensional carbon allotrope, fullerenes \cite{KrotoNat1985}, results in superconductivity \cite{HaddonNat1991,HebardNat1991,DahlkeJACS2000} with a record $T_{\text{c}}$ of $29$ K for Rb$_3$C$_{60}$ \cite{JanossyPRL1993}. Besides, alkali doped fullerenes form conducting polymers (e.g. KC$_{60}$ \cite{StephensNat1994,BommeliPRB1995,JanossyPRL1997}) with a density-wave correlated ground state. Alkali and alkali earth atom doped graphite \cite{DresselhausAIP2002} also give rise to superconductivity with the record $T_{\text{c}}$ of $11.5$ K for CaC$_6$ \cite{WellerNatPhys2005,EmeryPRL2005}. Similarly, alkali atom doped SWCNTs \cite{RaoNat1997} are compelling as these allowed the observation of a dimensional crossover from a 1D correlated metal, a so-called Tomonaga--Luttinger liquid \cite{IshiiNat2003} to a Fermi liquid \cite{RaufPRL2004}.

Graphene \cite{NovoselovSci2004}, the latest addition to the family of electron acceptor allotropes of carbon, can be also well doped with alkali atoms including Li, K, Rb \cite{JungACSNano2011,AjayanACSNano2011,HowardPRB2011,ParretACSNano2013}. We recently reported the successful synthesis of Na doped graphene \cite{MarkusACSNano2020}, which is surprising as Na does not intercalate the host compound, graphite to lower stages \cite{AsherNat1958,AsherJINC1959}. 

Besides the discovery of several fundamentally interesting and compelling phases, the technological importance of alkali atom doped carbon is unquestionable, as shown by e.g. the ubiquitous use of lithium batteries for whose development the Nobel prize in chemistry in 2019 was awarded.

The starting step of the intercalation has often been a gradual exposure to the vapor of the heavier alkali atoms (K, Rb, and Cs), especially when the stoichiometric content of the alkali/carbon ratio was unknown while monitoring the DC resistance of the respective materials \cite{HaddonNat1991,HebardNat1991,AkersTSF1995,PichlerPRL2001}. For graphene, measurement of DC resistance could be performed for small, individual flakes, however this method is clearly impractical for the study of the bulk form of graphene, after chemical exfoliation. 

Here, we report the systematic intercalation of chemically exfoliated graphene with potassium while monitoring the sample resistance using microwave conductivity, which is a contactless transport method and is readily applicable for air-sensitive, porous materials, where the conventional contact methods are not possible. We find that the conductivity of the samples increases by about an order of magnitude upon potassium intercalation. The increased number of charge carriers is deduced from the ESR intensity, which is also enlarged by the same amount, compared to the first doping step. In ESR two asymmetric Dysonian signals are found denoting metallicity. The width of the broader component grows with the doping steps, however, the narrow components seem to have a constant line width.

\section{Methods and sample preparation}

Few-layer graphene samples were prepared from saturation potassium-doped spherical graphite powder (SGN18, Future Carbon) using DMSO solvent for the wet chemical exfoliation as described elsewhere \cite{EnglertNC2011,VeceraPSSB2014,VeceraNatCom2017}. Chemical exfoliation was finalized using ultrasound tip sonication, as it is known to produce the best quality \cite{MarkusPSSB2015,SzirmaiSR2019}. The properties of the starting material are well characterized by atomic force microscopy and Raman spectroscopy, which revealed that restacked few-layer graphene is also present in the sample \cite{SzirmaiSR2019}. The dominant portion ($90\%$) of the material consists of $5$ layers or less, including monolayer content. Prior to \emph{in situ} measurements, the undoped FLG was heated to $400~^{\circ}$C for $30$ minutes in high vacuum ($2 \times 10^{-6}$ mbar) to remove any residual solvents. It was shown previously in Refs. \cite{MarkusPSSB2015} and \cite{SzirmaiSR2019} that this does not affect the morphology of the starting FLG.

Afterwards, the FLG material was placed in one end of a quartz tube, which is narrowed down in the middle. In the other end excess amount of potassium is placed (Aldrich MKBL0124V $99.9+$$\%$ purity). The geometry is similar to the one used in the two-zone vapor phase intercalation technique \cite{DresselhausAIP2002}. The ampoule is sealed under high vacuum ($2 \times 10^{-6}$ mbar). Afterwards, it is inserted into the microwave conductivity measurement setup (described later), where the \emph{in situ} intercalation takes place. The process is driven by the thermal and chemical potential gradient present between the two ends of the sealed ampoule. Two samples were investigated with the technique, one with slow intercalation steps, and one where the intercalation carried out in a more rapid manner. These samples are referred to as "\emph{in situ} \#1" and "\emph{in situ} \#2". In the "\emph{in situ} \#1" cycle, the sample was heated to $200~^{\circ}$C then cooled down instantly in the first $5$ steps. In the latter steps, the sample was kept at $200~^{\circ}$C for $30$ minutes, then cooled down to room temperature. Each point of the first $7$ steps consists of an average of $64$ rapid frequency sweep experiments. After the $7$th step, the number of averages was increased to $256$ to eliminate the higher noise arising from lower $Q$-factors, hence the overall signal-to-noise ratio is increased. In the case of "\emph{in situ} \#2" all the steps followed the latter protocol thus the doping proceeded faster, more aggressively. The conductivity of the samples was monitored continuously. The sample labeled as "furnace" was intercalated in a furnace. The center of the furnace was heated to $250~^{\circ}$C, where the potassium was located. Here, each intercalation step took $30$ minutes. After each step, for every sample, an ESR measurement was performed to monitor the amount of conducting electrons in the system.

Due to lack of substrate, significant restacking of individual graphene flakes occurs, leading to misaligned layers in the sample. The powder sample consists of sponge-like structures, as thermodynamic equilibrium is achieved to create a three-dimensional solid material.

Microwave conductivity measurements were performed with the cavity perturbation technique \cite{KleinIJIMW1993,DonovanIJIMW1993} in a custom-built high-temperature setup. The method is proven to give meaningful information about the conductivity of porous and air-sensitive samples \cite{MarkusPSSB2018} and well suited to study \emph{in situ} changes. The used copper cavity has an unloaded quality factor of $Q_0=10,000$ and a resonance frequency, $f_0 \approx 10.2$ GHz, whose temperature dependence is taken into account. The samples were placed in the node of the microwave electric field and maximum of the microwave magnetic field inside the TE$011$ cavity, which is the appropriate geometry to study minute changes in the conductivity \cite{KitanoPRL2002}. The alternating microwave magnetic field induces eddy currents in the sample, which causes a change in the microwave loss and shifts the resonator frequency. The $Q$-factor of the cavity is measured via rapid frequency sweeps near the resonance. A fit to the obtained resonance curve yields the position, $f$, and width, $\Gamma_f$, of the resonance. $Q$ is afterwards obtained from its definition $Q = f/\Gamma_f$. This value has to be corrected with the unloaded $Q$ factor of the cavity, thus the loss caused by the inserted sample is:
\begin{equation}
    \mathrm{\Delta} \left( \frac{1}{2Q} \right) = \frac{1}{2Q} - \frac{1}{2Q_0},
\end{equation}
where $Q_0$ is the $Q$-factor of the unloaded cavity.

Since the FLG material is known to have low conductivity compared to metals, the $Q \sim \varrho$ relation is used \cite{Csosz2018,MarkusPSSB2018}, where $\varrho$ is the resistivity of the sample.

Room temperature ESR measurements were performed after each doping step. The measurements were performed on a Bruker Elexsys E500 X-band spectrometer, without opening the sample ampoules. The spectral parameters (intensity and line width) of each signal component are determined by fitting (derivative) Lorentzian and Dysonian curves \cite{DysonPR1955}, as is customary in the ESR literature.

\section{\emph{In situ} microwave conductivity measurements}

\begin{figure}[h!]
    \includegraphics*[width=\linewidth]{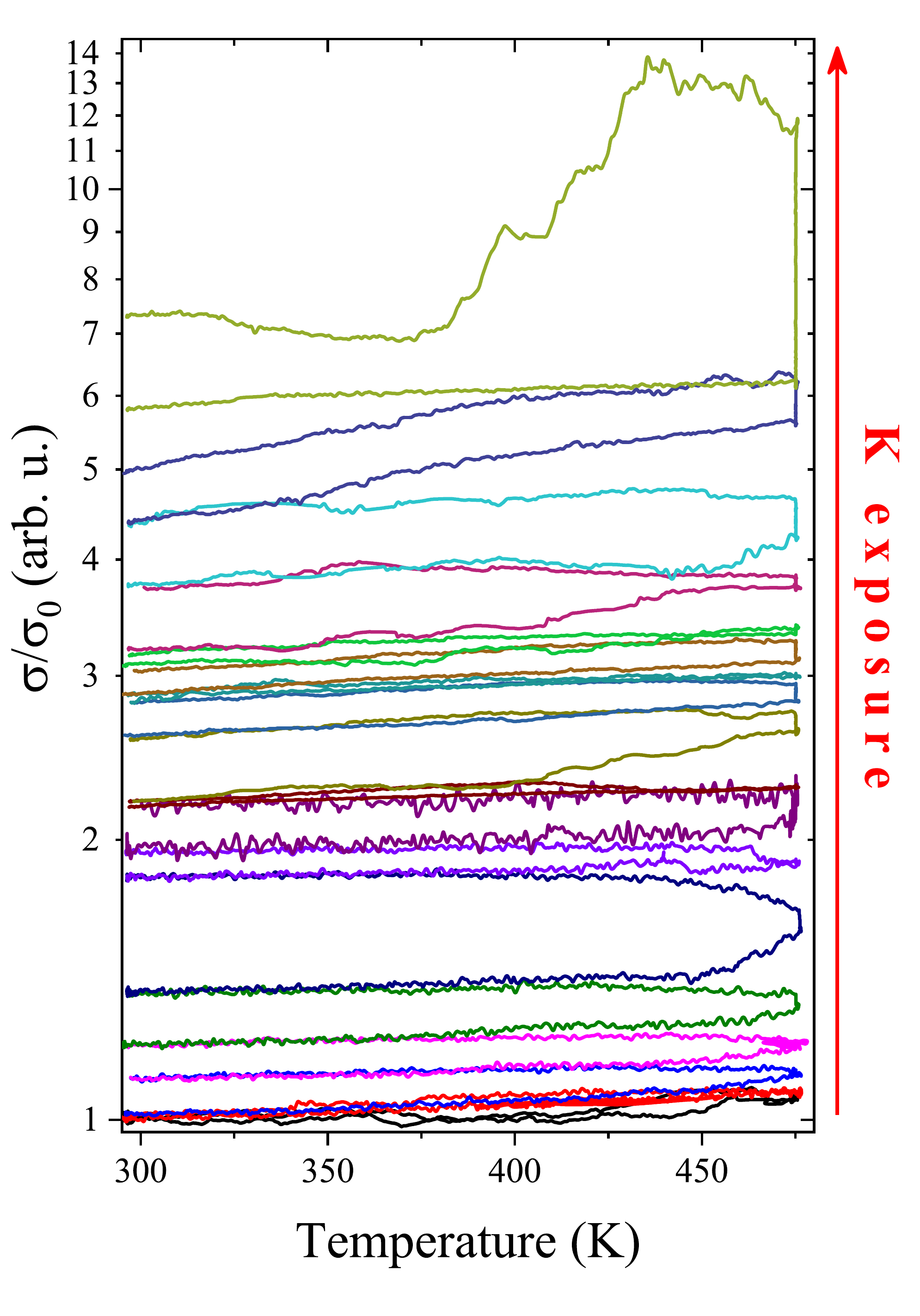}
    \caption{Microwave conductivity data normalized to the value of the undoped sample taken at room temperature. Upon potassium exposure, the conductivity increases in each step. In the first $5$ steps the ampoule was heated up to $200~^{\circ}$C then cooled down to room temperature, while in the latter steps the sample was kept at $200~^{\circ}$C for $30$ minutes. Note the gradual increase in the conductivity; after the final doping step, the value is increased by a factor of about $7.3$. The lower noise starting from step $8$ is due to the increased number of averages for each measurement point.}
    \label{fig:insitu_mwcond}
\end{figure}

Microwave conductivity results measured on the sample labeled as "\emph{in situ} \#1" are presented in Fig. \ref{fig:insitu_mwcond}.

The undoped material exhibit a flat, semi-conducting characteristic, reproducing our earlier results in Ref. \cite{MarkusPSSB2015}. This behavior is due to the low amount of mobile charge carriers present in the material; most of the conducting electrons are thermally excited. The diffusion of the potassium atoms becomes dominant above $175~^{\circ}$C and causes a gradual upturn in each curve. The increased conductivity is a clear sign that the material becomes intercalated and the few-layer graphene accommodates the potassium ions and its electron is donated to the host material. After each step, the ampoule was pulled out from the setup and it was ensured that no metallic potassium was left on the inner wall of the quartz tube. Then an ESR measurement was performed (next section). The doping proceeds through several steps with monotonously increasing conductivity. In the final step, we observe an overall increase of the conductivity by a factor of $7.3$ at room temperature. Even though the resistivity of the final product is about one order smaller compared to the starting material it still presents a semi-conducting behavior. This is either due to an incomplete charge transfer (in several cases vapor-phase intercalation cannot reach very high doping, see e.g. potassium doped single-walled carbon nanotubes \cite{SzirmaiPRB2017}), or the simple relation of $Q\sim \varrho$ does not hold anymore. The latter can occur if the flakes develop electronic contacts among each other and the whole sample starts to behave as a bulk metal instead of a loosely packed powder. In this scenario, the microwaves can only penetrate to a finite depth due to the skin effect causing $Q\sim \sqrt{\sigma}$, which would yield a metallic behavior \cite{Csosz2018}. Unfortunately, investigating such a turnover is beyond the scope of the present manuscript. Furthermore, this effect can also be caused by the freezing-out of the diffusing potassium atoms as the temperature is lowered. 

\section{Spin properties}

\begin{figure}[h!]
    \includegraphics*[width=\linewidth]{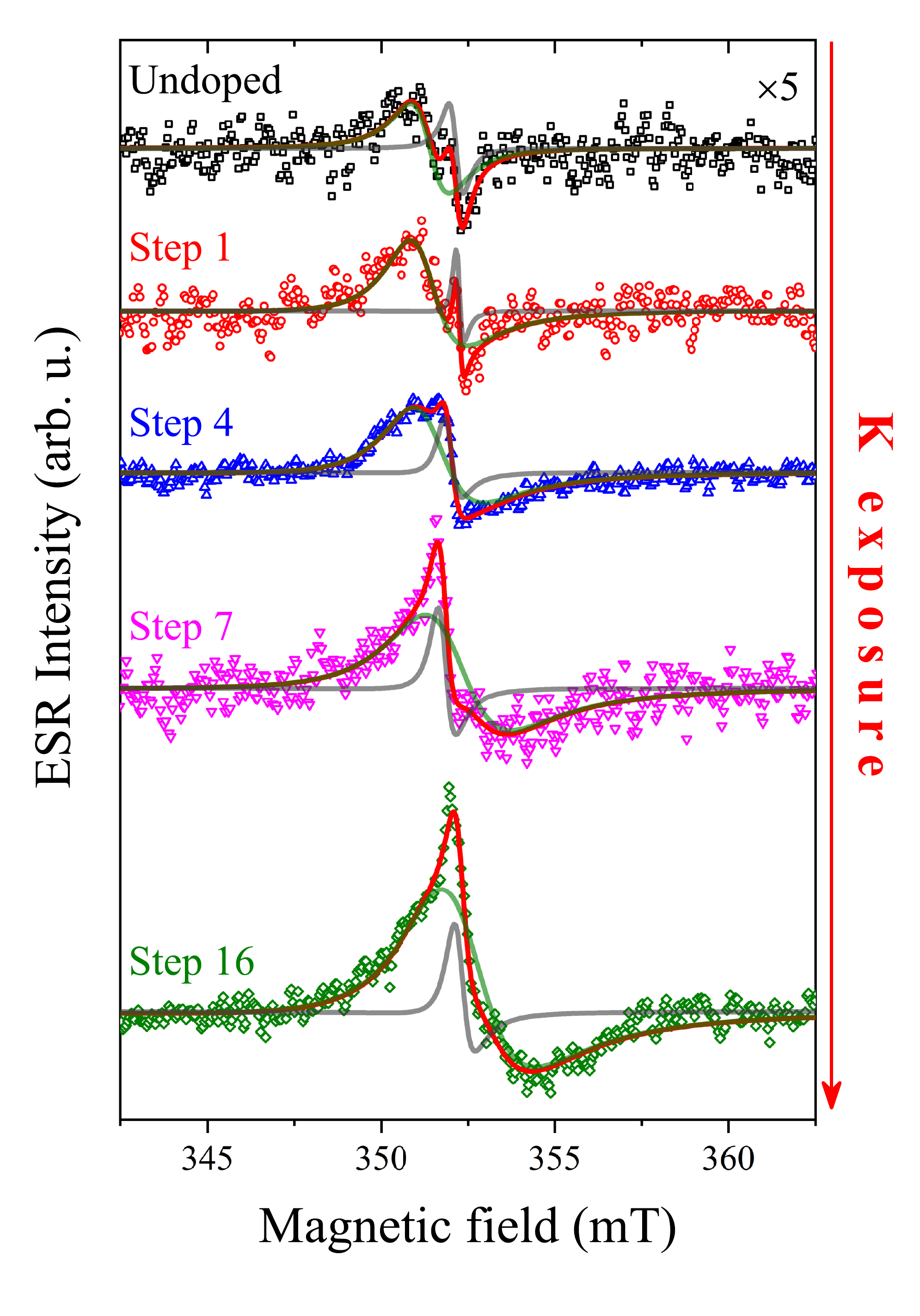}
    \caption{X-band ESR spectra of the undoped material and after the $1$st, $4$th, $7$th, and $16$th doping steps. The undoped FLG has two, symmetric components due to the presence of localized defects, as seen before \cite{MarkusPSSB2015,MarkusACSNano2020}. After potassium intercalation, two, different asymmetric lines appear. These lines are identified as Dysonians \cite{DysonPR1955} as an indication of the presence of conducting electrons. The intensity and the asymmetry of the Dysonians develop monotonously with the amount of alkali in the material.}
    \label{fig:esr_spectr}
\end{figure}

ESR spectra of the undoped material and after the $1$st, $4$th, $7$th, and $16$th doping steps are presented in Fig. \ref{fig:esr_spectr}.

The undoped material exhibits two symmetric Lorentzian peaks, as noted earlier \cite{MarkusPSSB2015,MarkusACSNano2020}. The presence of these signals is due to the remaining solvent, dangling bonds, or other localized lattice defects. Even after the first intercalation step, two, different Dysonian features \cite{DysonPR1955} appear with a gradually higher intensity. The presence of such a line shape is an indication for the appearance of new, conducting electrons in the system. The asymmetry and the intensity of the Dysonians grow with the amount of alkali in the system. Previously, in the Li-doped FLG system, we also observed two Dysonians, however, the sodium system only exhibits a single Pauli-type signal \cite{MarkusACSNano2020} as Na cannot penetrate into graphite, it only intercalates the monolayer flakes. Here, similarly to the lithium-doped FLG system, the two lines can be correlated to the simultaneous presence of flakes with single-layer together with multilayer flakes.

\subsection{Increased number of charge carriers as seen by ESR}

\begin{figure}[h!]
    \includegraphics*[width=\linewidth]{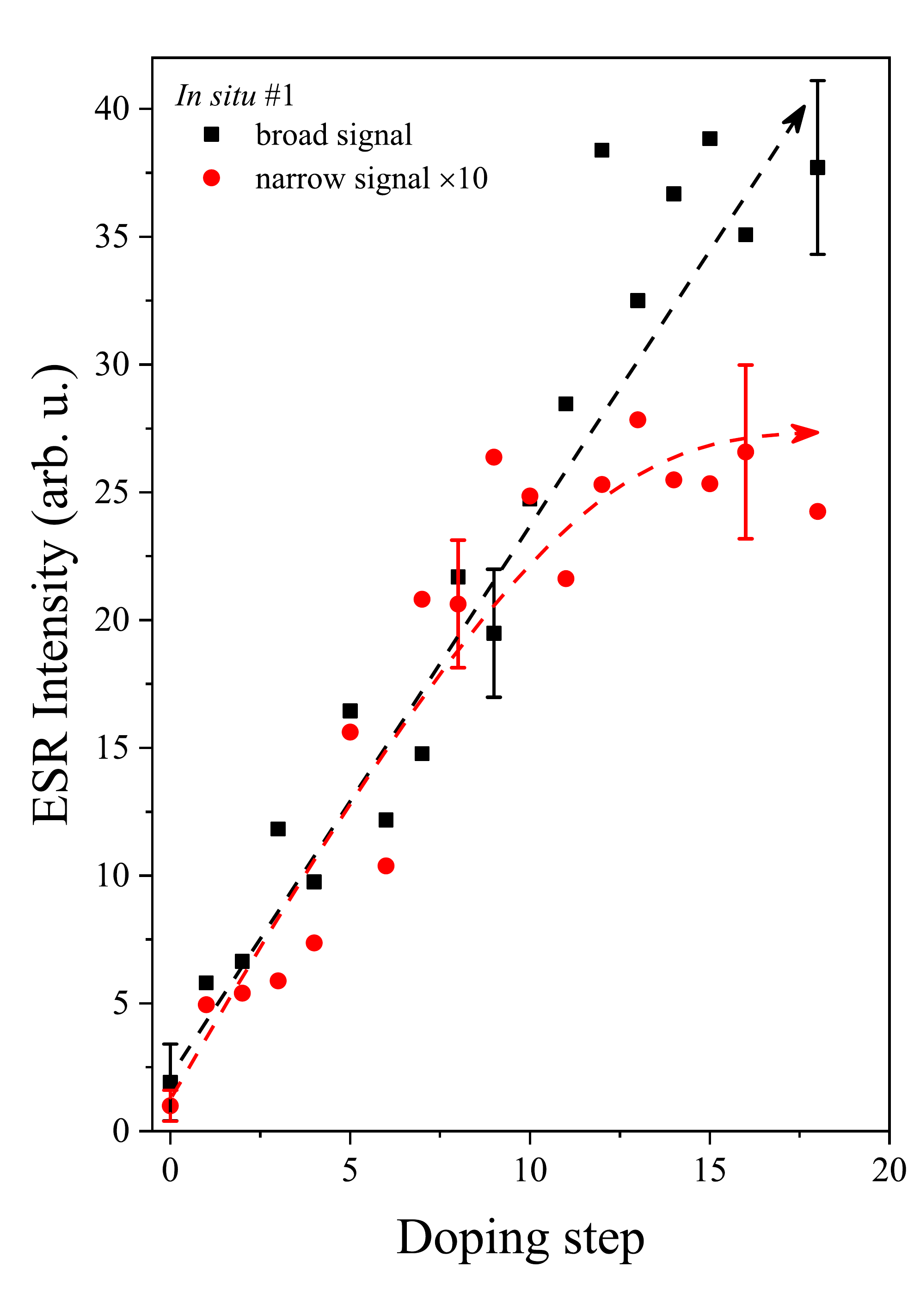}
    \caption{Fitted intensities of the observed ESR lines as a function of the doping steps for the sample labeled as "\emph{in situ} \#1". The values of the narrow line are multiplied by $10$ to be scaled with the broader component. Please note the gradual increase for both spin species, the broader component is amplified by a factor of about $8$, while the narrow is by $5$, compared to the $1$st step. The dashed arrows are denoting the general trends observed in the material.}
    \label{fig:esr_int}
\end{figure}

The ESR intensities found by fitting the two signals are presented in Fig. \ref{fig:esr_int}. The intensity of the narrow signal is multiplied by $10$ to be scaled with the broader component.

The intensity, which is directly proportional to the spin-susceptibility, of both Dysonian components, increases with the amount of potassium. This indicates that after each intercalation step more and more charge is transferred to the FLG host. The intensity of the broader component is grown by a factor $8$, while the narrow component is increased by a factor of about $5$ compared to the $1$st doping step. Surprisingly, the narrow signal seems to saturate after the $10$th step which is probably related to the special, mixed structure of the material (simultaneous presence of single- and multilayers).

\subsection{Tuning spin properties}

\begin{figure}[h!]
    \includegraphics*[width=\linewidth]{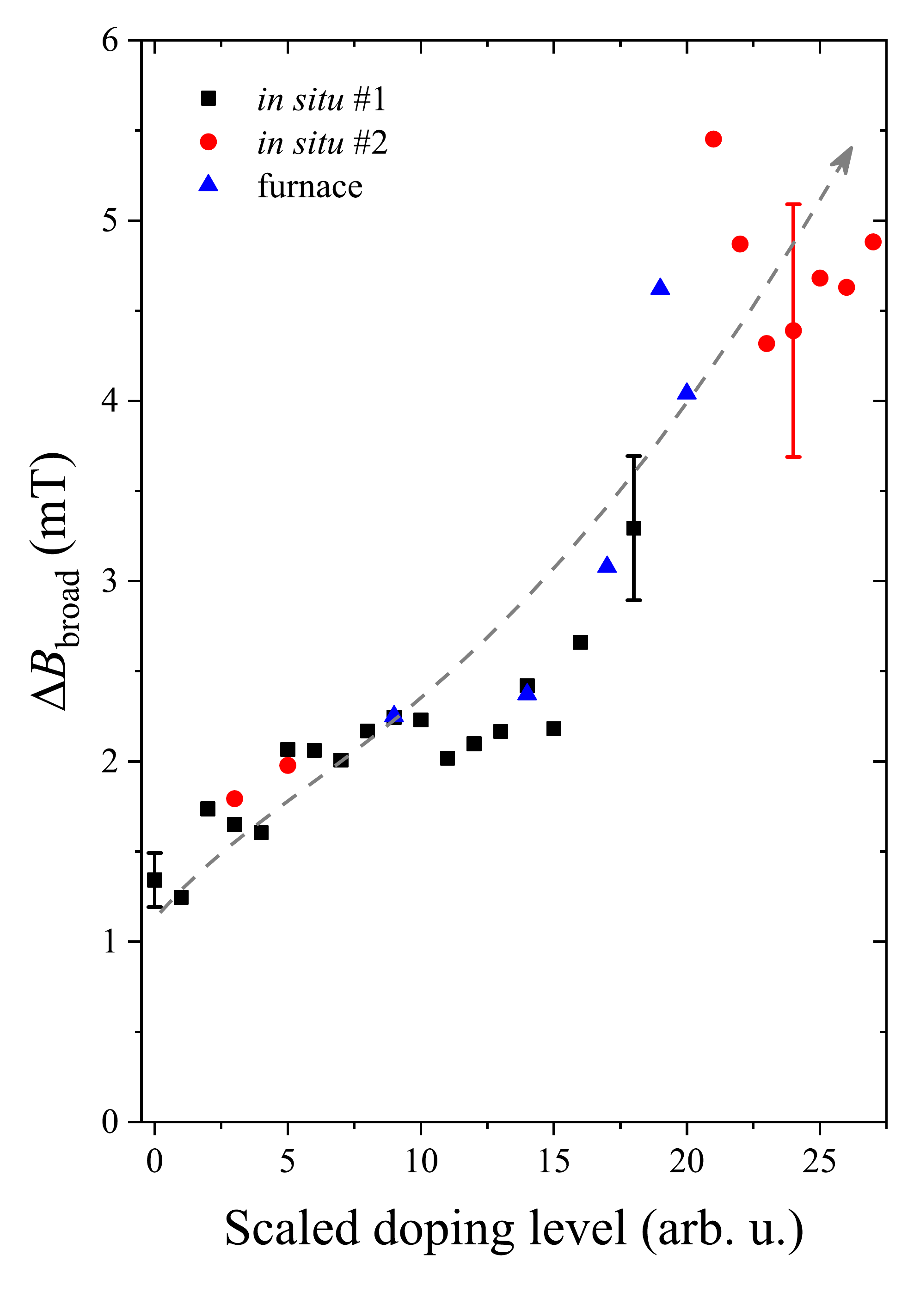}
    \caption{ESR line width of the broader component as a function of scaled doping level. As the number of mobile charge carriers is increased the line width is gradually increased and the spin relaxation time is reduced. The dashed arrow is denoting the general trend observed in the material.}
    \label{fig:esr_broad}
\end{figure}

\begin{figure}[h!]
    \includegraphics*[width=\linewidth]{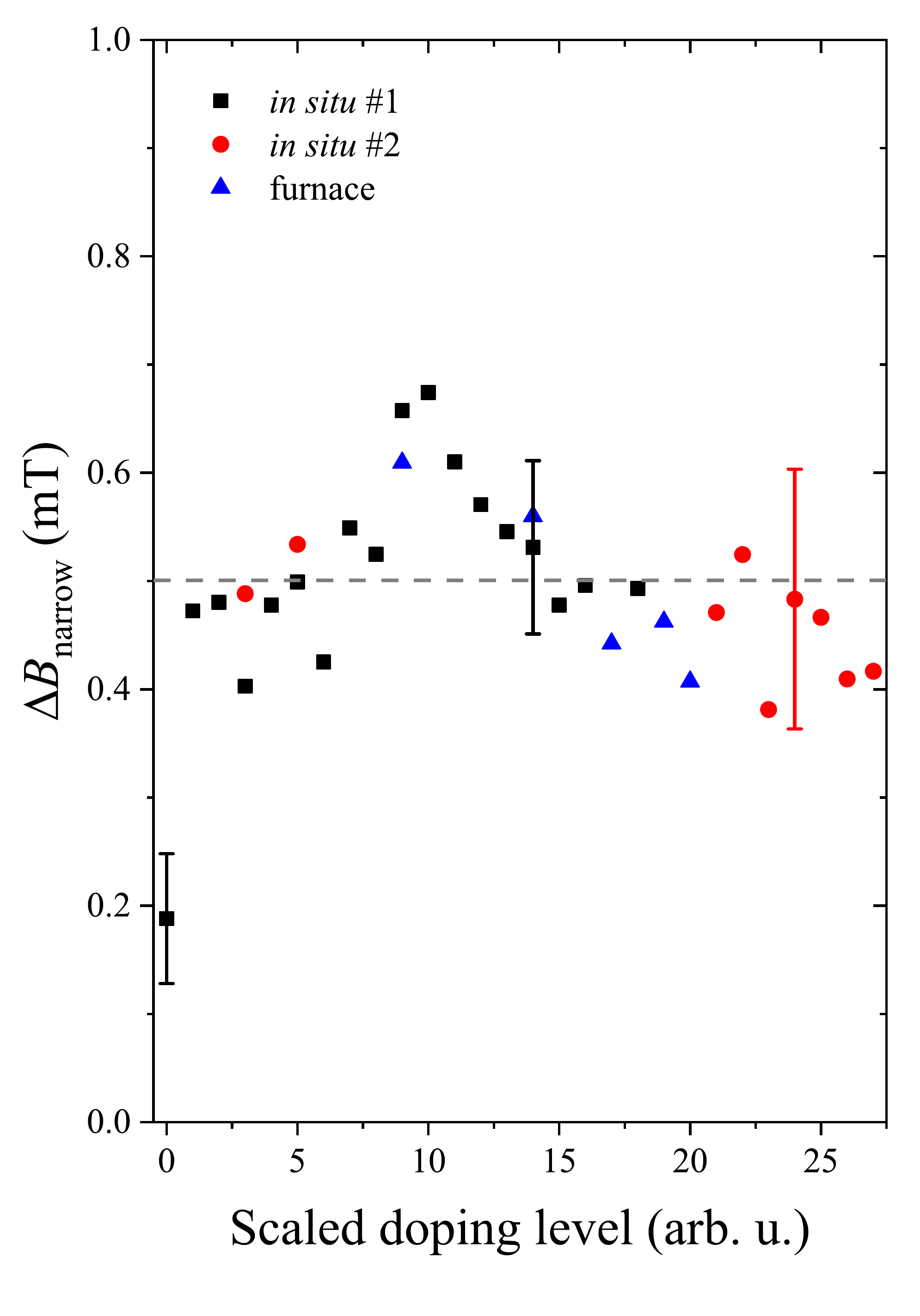}
    \caption{ESR line width of the narrower component as a function of scaled doping level. This component does not seem to show any trend with the doping steps. Instead, it settles around the value of $0.5$ mT (dashed grey line). This yields a spin relaxation time of $12$ ns on average.}
    \label{fig:esr_narrow}
\end{figure}

The evolution of ESR line widths is depicted in Figs. \ref{fig:esr_broad} and \ref{fig:esr_narrow} for the broad and narrow component, respectively. The scaled doping level is obtained by arranging the widths of the broader component. The values are not interchanged, only shifted to achieve an increasing order among the different samples. This can be done since the line width is unaffected by the variations in the sample masses and geometry. As the charge transfer is a monotonous function of the doping steps carried out, as concluded from the conductivity data, the same monotonous fashion is assumed for the line width of the broader component. This is also confirmed by stand-alone experiments that show this behavior. The breadth of the narrow component is treated together with the broad line and thus sorted accordingly.

The breadth of the broader component is increasing with the amount of charge present in the system. The corresponding spin relaxation time in the first few steps is about $3.5$ ns and it goes down to $1.2$ ns. The observed values are smaller compared to Stage-I, II, and III compounds formed with potassium in graphite \cite{DelhaesMSE1977,LauginPBC1980}. However, the origin of this behavior is yet unclear.

The width of the narrow line does not develop significantly upon potassium intercalation. In stark contrast, it settles around the value of $0.5$ mT after the first step. The calculated spin relaxation value is $12(1)$ ns. Compared to graphite intercalation compounds (GICs) this value is close to the $13$ ns observed in the Stage-III potassium compound \cite{LauginPBC1980}. However, a Stage-III material would not exist through all the doping steps.

\vspace{5mm}

\section{Summary}

The present work presents \emph{in situ} conductivity measurements upon potassium intercalation. The evolution of resistivity is followed thoroughly and we found that it is lowered by a factor of $7.3$ compared to the undoped material at room temperature. After each doping step an ESR experiment is performed, where two Dysonian lines are identified in the doped material. The increased amount of charge carriers is noted, as suggested by the increase of the intensity. The width of the broader line monotonously increases with each doping step going from $16$ mT up to $48$ mT in the final step. Contrary, the narrow component has a fixed width of $0.5$ mT, which does not change with the amount of potassium present in the system. The calculated spin relaxation time for this component is around $12$ ns, which supposed to be large enough for spintronics applications.

\section{Acknowledgements}
Work supported by the Hungarian National Research, Development and Innovation Office (NKFIH) Grant Nr. K119442, and 2017-1.2.1-NKP-2017-00001. The research reported in this paper was supported by the BME-Nanonotechnology FIKP grant of EMMI (BME FIKP-NAT). P.~Sz, B.~N., and L.~F. were supported by the Swiss National Science Foundation (Grant No. 200021 144419). K.~F.~E., A.~H., and F.~H. thank the Deutsche Forschungsgemeinschaft (DFG-SFB 953, Synthetic Carbon Allotropes” Project A1) for financial support.

\bibliography{k-flg-insitu_mw-esr}



\end{document}